\documentclass[onecolumn,secnumarabic,amssymb, amsmath, nofootinbib,tightenlines,
nobibnotes, aps, prl,epsfig,pdffig]{revtex4}
\usepackage{graphicx}% Include figure files
\usepackage{dcolumn}% Align table columns on decimal point
\usepackage{bm}% bold math
\begin{document}
\preprint{APS/123-QED}
\title{An analysis of the fragmentation function of gluon at next-to-leading order
approximation}% Force line breaks with \\

\author{H.S.Nakhaei}%
 \email{ SaghaeNakhaei.Hossein@razi.ac.ir}
\author{G.R.Boroun}%
 \email{ boroun@razi.ac.ir}
\affiliation{ Department of physics, Razi University, Kermanshah
67149, Iran}% \textbackslash\textbackslash

\begin{abstract}
%%%%%%%%%%%%%%%%%%%%%%%%%%%%%%%%%%%%%%%%%%%%%%%%%%%%%%%
We are investigating the behavior of the fragmentation function of
a gluon, denoted as $ D_{g}(x,\mu^2)$, where $\mu$ represents the
observable scale. This function is derived from the
Dokshitzer-Gribov-Lipatov-Altarelli-Parisi (DGLAP) evolution
equations. Our objective is to evolve the fragmentation function
of a gluon for heavy-quark-antiquark bound states with large
transverse momentum using a Laplace transform technique. This
method enables us to calculate numerical solutions for two and
four quarkonium states based on the known initial fragmentation
function of gluons. We examine both leading-order (LO) and
higher-order approximations for the fragmentation function of a
gluon, $[g{\rightarrow}T_{nQ}]$, by integrating the evolved
fragmentation function of the gluon at the initial scale. In our
computations, we utilize the initial scales for $T^{2c}_{g}$ from
Braaten-Yuan [E.Braaten and T.C.Yuan, Phys. Rev. Lett. {\bf71},
1673 (1993)] and for $T^{4c}_{g}$ and $T^{4b}_{g}$ from
Celiberto-Gatto-papa [ F. G. Celiberto, G. Gatto and A. Papa, Eur.
Phys. J. C {\bf84}, 1071 (2024)] and Celiberto-Gatto [ F. G.
Celiberto and G. Gatto, Phys. Rev. D {\bf111}, 034037 (2025)]
respectively. Through comparing our predictions with existing
literature results, we can accurately determine the evolution of
the fragmentation function
of a gluon at scale $\mu$.\\

%%%%%%%%%%%%%%%%%%%%%%%%%%%%%%%%%%%%%%%%%%%%%%%%%%%%%%%
\end{abstract}
 \pacs{***}%PACS, the Physics and Astronomy
                              %Classification Scheme.
 %Use showkeys class option if keyword
                              %display desired
\maketitle
%**********************************************************
%%%%%%%%%%%%%%%%%%%%%%%%%%%%%%%%%%%%%%%%%%%%%%%%%%%%%%%%%%%%%%%%%%%%%%%%%%%%%%%%%%%%%%%%%%%%%
\subsection{I. Introduction}

The theory that describes the strong interaction, in which the
fundamental degrees of freedom are quarks and gluons, is Quantum
Chromodynamics (QCD). QCD, as the fundamental theory, describes
the strong interactions of heavy quarkonia with large transverse
momentum $P_{T}$. The dominant mechanism for the production of a
parton with high transverse momentum is fragmentation, which
subsequently decays into the quarkonium state and other partons
\cite{R1}. A measurement of gluon fragmentation in various three-
jet topologies has been performed in the DELPHI experiment
\cite{R2}. The scaling violation for gluon jets appears to be
significantly stronger than for quark fragmentation. The
probability for a virtual gluon to decay into a $\psi$ was
calculated in Ref.\cite{R3}, where they studied $J/\psi$
production from gluon jets at the CERN $e^{+}e^{-}$ collider, LEP.
In Ref.\cite{R4}, a correct definition of the gluon FF,
$D^{H}_{g}(z)$, is presented starting from the single gluon
incoming state $|g
>$  with  $Q^{a}_{\mu}(x)$, which is
the quantum gluon field that fragments into hadron $H$. The
authors demonstrate that using the non-abelian field
tensor\footnote{ Based on the Yang-Mills construction, the
coupling between the gluonic field and matter is defined through
the covariant derivative
$D_{\mu}=\partial_{\mu}-igT^{a}Q^{a}_{\mu}$ with
$T^{(A)a}_{bc}=-if^{abc}$.}
$F^{a}_{\mu{\upsilon}}(x)={\partial}_{\mu}Q^{a}_{\upsilon}(x)-{\partial}_{\upsilon}Q^{a}_{\mu}(x)
+gf^{abc}Q^{b}_{\mu}(x)Q^{c}_{\upsilon}(x)$ instead of
$Q^{a}_{\mu}(x)$ in the initial state is not in line with the
factorization theorem in QCD. They conclude that the accurate
definition of the gluon fragmentation function at high energy
colliders is derived from the single gluon initial state $|g
>$ , is gauge invariant, and aligns with the factorization theorem
in QCD at high-energy colliders.\\
The fragmentation functions (FFs) $D^{H}_{g,q}(x,\mu)$ into a
hadron are studied for individual gluon and quark jets in three
jet events originating from $Z$ decays using the DGLAP equations
\cite{R5, R6, R7} by evolving from the initial FF
$D^{H}_{g,q}(x,\mu_{0})$. The FFs can be defined as the scattering
cross section of the process $A+B{\rightarrow}H+\mathrm{Jet}$:
\begin{eqnarray}
d\sigma=\sum_{a,b,c}\int_{0}^{1}dx_{a}\int_{0}^{1}dx_{b}\int_{0}^{1}dz
f_{a/A}(x_{a},\mu)f_{b/A}(x_{b},\mu)d\widehat{\sigma}(a+b{\rightarrow}c+\mathrm{Jet})
D_{H/c}(z,\mu),
\end{eqnarray}
where $a$ and $b$ are incident partons in the colliding hadrons
and $f_{i/A}(i=a,b)$ are the parton distribution functions at the
scale $\mu$. The FFs at the initial scale can be find in
literature \cite{R1, R8, R9, R10}. In Ref.\cite{R1}, the authors
calculated\footnote{In Feynman gauge, the FF for
$g^{*}{\rightarrow}\eta_{c}g$ at the initial scale reads
 $$
D_{g{\rightarrow}\eta_{c}}(z,2m_{c})=\frac{1}{24\pi}\alpha_{s}(2m_{c})^{2}\frac{|R_{(0)}|^{2}}{m_{c}^{3}}[3z-2z^2+2(1-z){\ln}(1-z)],
  $$
  where the value of the $S$-state wave function
at the origin $R(0)$ is determined from the $\psi$ electronic
width to be $|R(0)|^2 = (0.8~\mathrm{GeV})^3$ with
$\alpha_{s}(2m_{c})=0.26$ which $m_{c}=1.5~\mathrm{GeV}$
\cite{R1}.} the FF $D_{g{\rightarrow}\eta_{c}}(z,2m_{c})$ for a
gluon to split into the $^{1}S_{0}$ charmonium state $\eta_{c}$ to
leading order in $\alpha_{s}(2m_{c})$. In Refs.\cite{R8, R9}, the
authors defined the FFs at the initial scale, focusing on the
fragmentation of a gluon into the $S$-wave spin singlet and
spin-triplet heavy quarkonia considering the processes
$g{\rightarrow}\eta_{c}/\eta_{b}$ at the leading-order (LO) and
next-to-leading order (NLO) approximations\footnote{At the LO
approximation, FF reads
$$
D^{LO}_{g{\rightarrow}\eta_{i}}(z,2m_{i})=\frac{Nz}{F(z,<p^{2}_{T}>)}\bigg{[}
2m_{i}+\frac{z<p^{2}_{T}>}{M(1-z)}+\frac{M(1-z)}{z}\bigg{]},~~i=c,b
$$
where $N=\frac{2}{9}(\pi{m_{c}^{7/2}}f^{2}_{M}C_{F}\alpha_{s})$
and
$$
F(z,<p^{2}_{T}>)=\bigg{[}\frac{z^2<p^{2}_{T}>+M^2(1-z)^2}{Mz(1-z)}(M^2
+\frac{z^2<p^{2}_{T}>}{1-z})\bigg{]}^2,
$$
where $M$ is the meson mass and
$f_{M}=\sqrt{\frac{12}{M}}|\psi(0)|$ with
$m_{c}=1.67~\mathrm{GeV}$
 and $m_{b}=4.78~\mathrm{GeV}$.\\
 At NLO approximation, the FF reads
$$
D^{NLO}_{g{\rightarrow}\eta_{i}}(z,2m_{i})=N(D^{LO}_{g{\rightarrow}\eta_{i}}(z,2m_{i})+D^{Real}_{g{\rightarrow}\eta_{i}}(z,2m_{i}))
$$
where $N$ is the normalization constant under the condition
$\int_{0}^{1}D_{g}(z,2m_{i})dz=1$. The superscript
$^{"}\mathrm{Real}^{"}$ refers to the contribution of Feynman
diagrams including real gluon emissions \cite{R8}.} based on the
Suzuki$^,$s model\footnote{In the suzuki model, the FFs depend on
the transverse momentum $k_{T}$ of the initial parton, which is
treated as a phenomenological parameter.} \cite{R10}.\\
In Ref.\cite{R11}, relativistic correction effects to the
$g{\rightarrow}\eta_{c}$ FF in the  effective-field-theory
approach known as nonrelativistic QCD (NRQCD) factorization are
discussed. In this model, the FF of a gluon fragmenting into the
pseudoscalar quarkonium is expanded in terms of the characteristic
velocity $q(\overline{q})$ of the quark inside a charmonium
denoted as $v$, by the following form
\begin{eqnarray}
D_{g{\rightarrow}\eta_{i}}(z)=d^{(0)}(z)<\mathcal{O}_{1}^{\eta_{i}}>+d^{(2)}(z)\frac{<\mathcal{P}_{1}^{\eta_{i}}>}{m^2}+\mathcal{O}(v^3),~~i=c,b
\end{eqnarray}
Here, $d^{(0)}(z)$ and $d^{(2)}(z)$ are the corresponding
short-distance coefficient functions. $\mathcal{O}_{1}^{\eta_{i}}$
and $\mathcal{P}_{1}^{\eta_{i}}$ are color-singlet NRQCD
production operators\footnote{ For the pseudoscalar quarkonium
$\eta_{c}$, we have
$$
\mathcal{O}_{1}^{\eta_{c}}=2N_{c},~~~
\mathcal{P}_{1}^{\eta_{c}}=2N_{c}(E^2-m^2),
$$
where $N_{c}=3$. The corresponding short-distance coefficient
functions accurate through order-$v^2$, are defined at the LO
approximation by the following forms:
$$
d^{(0)}(z)=\frac{\alpha_{s}^2}{4N_{c}^2m^3}[3z-2z^2+2(1-z){\ln}(1-z)],~~~
d^{(2)}(z)=-\frac{11}{6}d^{(0)}(z).
$$
Additionally, in Ref.\cite{R11}, the $g{\rightarrow}\eta_{c}$ FF
modified from the inclusive $\eta_{c}$ production from Higgs boson
decay $h{\rightarrow}g^{*}g{\rightarrow}\eta_{c}+gg$ is defined.}.
In Refs.\cite{R12, R13, R14}, the evolution of the gluon FF in the
modified leading logarithm approximation (MLLA)\footnote{The MLLA
is a systematic improvement over an earlier approximation known as
the Double Logarithmic Approximation (DLA), which involves
resumming leading logarithms by summing tree -level diagrams where
outgoing gluons are strongly ordered in their angles of emission.}
is discussed. The authors present the evolution of the gluon FF at
the LO and higher-order approximations. They demonstrate that the
higher moments of the evolution obtained from the MLLA are
considered spurious higher -order terms in perturbation theory.\\
The FF of gluon into spin-singlet P-wave quarkonium was developed
by Collins and Soper using the velocity expansion in the NRQCD
factorization approach. This FF was also applied to $B_{c}^{(*)}$
within the NRQCD factorization framework at the lowest order in
velocity expansion and strong coupling constant. These
developments are discussed in Refs.\cite{R15} and \cite{R16}
respectively. Analytical solutions for DGLAP evolution equations
for FFs of mesons and protons based on the Laplace transform
method are discussed in Refs.\cite{R17, R18, R19}. The symmetry
breaking of the sea quarks Fragmentation Functions according to
their mass ratio is studied in Ref.\cite{R17}. The cross section
of meson production via quark fragmentation is discussed in
Refs.\cite{R17, R20}.\\
Recently, the inclusive production of fully charmed tetraquarks in
high-energy proton collisions has been considered in
Refs.\cite{R21, R22, R23}. The first exotic hadron, the $X(3872)$,
was observed in 2003 by Belle at KEKB \cite{R24}. Then, in 2020,
the LHCb experiment observed the $X(6900)$ from the $J/\psi$-pair
production in the mass region between $6.2$ and
$7.4~\mathrm{GeV}/c^2$, covering the masses of states composed of
four charm quarks \cite{R25}. It can be considered as a tetraquark
or pentaquark. In the lowest Fock state, the $T_{4c}$ tetraquark
state is represented as $|cc\overline{c}\overline{c}>$. The
production mechanisms are based on the leading-twist fragmentation
of a single parton into the observed hadron. The FF of a gluon to
$T_{2c}$ and $T_{4c}$ states within a potential-quark become
dominant over direct short-distance production when the transverse
momentum exceeds ${\lesssim}10~\mathrm{GeV}$ and $20~\mathrm{GeV}$
respectively \cite{R26}. The fully charmed tetraquark, $T_{4c}$,
is known to have quantum numbers $J^{PC}$ of  $0^{++}$ for  the
ground state and $2^{++}$ for the radial excitation. In
Ref.\cite{R22} the tetraquark collinear fragmentation functions of
$T_{4c}(0^{++})$ and $T_{4c}(2^{++})$ are analyzed for bound
states. This analysis focuses on single-parton collinear
fragmentation in a variable-flavor number scheme (VFNS). The gluon
fragmentation into a color-singlet S-wave $T_{4c}(J^{PC})$ for
only the lowest-order contributions in the constituent
quarks$^{,}$ relative velocity can be described by the equation:
\begin{eqnarray}
D_{g}^{T_{4c}(J^{PC})}(z,4m_{c})=\frac{1}{m_{c}^{9}}\sum_{[n]}\mathcal{D}_{g}^{(J^{PC})}
(z,[n])<\mathcal{O}^{T_{4c}(J^{PC})}([n])>.
\end{eqnarray}
Here $m_{c}=1.5~\mathrm{GeV}$, and $\mathcal{D}_{g}^{(J^{PC})}
(z,[n])$ represents the short-distance coefficient (SDC) for the
$[g{\rightarrow}(cc\overline{c}\overline{c})]$ perturbative
transition.  $<\mathcal{O}^{T_{4c}(J^{PC})}([n])>$ denotes the
color-composite of the long-distance matrix elements (LDMEs) that
illustrate the nonperturbative hadronization of the
$T_{4c}(J^{PC})$ state. The quantity of
$\mathcal{D}_{g}^{(J^{PC})} (z,[n])$ for the composite quantum
number $[n]$ is determined using the color diquark-antidiquark
basis to decompose a color-singlet tetraquark state. The quantity
 $<\mathcal{O}^{T_{4c}(J^{PC})}([n])>$ represents the nonperturbative contribution to the FF at the initial
 scale. This is obtained by calculating the radial wave functions at
  the origin using potential models. In Ref.\cite{R23}, bottomoniumlike states in proton
  collisionsin the semi-inclusive hadroproduction of doubly
bottomed tetraquarks $(X_{b\overline{b}q\overline{q}})$ as well as
fully bottomed ones $(T_{4b})$ are discussed. The explicit form of
the $g{\rightarrow}T_{4b}$ at the initial scale $\mu=4m_{b}$ with
$m_{b}=4.9~\mathrm{GeV}$ is the same as for $g{\rightarrow}T_{4c}$
(i.e., Appendix A). The values of the LDMEs of $T_{4b}(0^{++})$
and $T_{4b}(2^{++})$ fully bottomed states evaluated within the
Coulomb potential diquark model are defined by the following form
\cite{R23, R26}
\begin{eqnarray}
<\mathcal{O}^{T_{4b}(J^{PC})}([n])>{\approx}400<\mathcal{O}^{T_{4c}(J^{PC})}([n])>.
\end{eqnarray}
The functions $\mathcal{D}_{g}^{(J^{PC})} (z,[n])$ and the values
of $<\mathcal{O}^{T_{4c}(J^{PC})}([n])>$
are presented in Appendix A.\\
In this article, we investigate the behavior of the fragmentation
function of a gluon that addresses the inclusive production in
high-energy proton scattering at the bound states $T_{2c}$,
$T_{4c}$ and $T_{4b}$ into the fragmentation function of a gluon
at the initial scale using Laplace transform techniques. We
utilize the Laplace transform technique to solve the DGLAP
evolution equation for $D_{g}(z,\mu)$ by employing the
$D_{g}(z,\mu_{0})$ at leading order (LO) and next-to-leading
order (NLO) approximations.\\
In the next section, we will summarize the Laplace transform used
to obtain an analytical solution for the fragmentation function of
a gluon at both the LO and NLO approximations. Following that, in
Sec. III, we will present the numerical results and compare them
with the existing FFs for the $[g{\rightarrow}T_{nQ}]$
fragmentation mechanism. The conclusions will be provided in Sec.
IV. Additional information on fragmentation functions at the
initial scales for the fully charmed tetraquarks and splitting
functions in $s$-space can be found in the Appendix.\\

\subsection{II. Fragmentation Functions}

The production of hadrons is not a perturbative process;
therefore, we cannot compute the FFs from perturbation theory. We
consider the change in FF $D_{i}(z,\mu)$ when the scale is
increased from $\mu$ to $\mu+d\mu$ by the DGLAP equations:
\begin{eqnarray}
\dfrac{{\partial}D^{T_{nQ}}_{i}(z,\mu)}{{\partial}\ln(\mu)}=\sum_{j}D^{T_{nQ}}_{j}(z,\mu)\otimes
P_{ji}(z,a_{s}(\mu)),~~~i,j=g,q,
\end{eqnarray}
where
$D_{s}(z,\mu^{2})=\sum_{i}(D_{q_{i}}+D_{\bar{q}_{i}})(z,\mu)$ and
$D_{g}(z,\mu)$ are the singlet and gluon fragmentation functions
respectively. The symbol $\otimes$ denotes convolution according
to the usual prescription. In a FF, $z$ represents the fraction of
a parton momentum carried by a produced parton. The splitting
functions $P_{ji}$ can be computed perturbatively in the strong
coupling constant:
\begin{eqnarray}
P_{ji}(z,a_{s}(\mu))=\sum_{k=0}a_{s}^{k+1}P^{(k)}_{ji}(z),~a_{s}=\frac{\alpha_{s}}{4\pi},
\end{eqnarray}
where the lowest-order functions $P^{(0)}_{ji}(z)$ are the same as
those in deep inelastic scattering (DIS) but the higher-order
terms are different \cite{R27} (please refer to Appendix B). The
FF of the gluon can be considered from the gluon-to-gluon
splitting only via the DGLAP equation as
\begin{eqnarray}
\dfrac{{\partial}D^{T_{nQ}}_{g}(z,\mu)}{{\partial}\ln(\mu)}=D^{T_{nQ}}_{g}(z,\mu){\otimes}\bigg{[}a_{s}P^{(0)}_{gg}(z,\mu)+a^{2}_{s}P^{(1)}_{gg}(z,\mu)\bigg{]}
 ,
\end{eqnarray}
where the splitting functions at the LO and NLO approximations are
defined in appendix B.\\
In the following, we will utilize the method developed in detail
in \cite{Block1, Block2,Block3,Block4, BH1, BR1} to determine the
FF of gluon using a Laplace-transform method. We now express the
DGLAP evolution equation for the FF of gluon [i.e., Eq. (7)] in
terms of the variables $\upsilon={\ln}(1/z)$ and $\mu$ along with
the notation
$\widehat{D}^{T_{nQ}}_{g}(\upsilon,\mu){\equiv}{D}^{T_{nQ}}_{g}(e^{-\upsilon},\mu)$,
as follows:
\begin{eqnarray}
\dfrac{{\partial}\widehat{D}^{T_{nQ}}_{g}(\upsilon,\mu)}{{\partial}\ln(\mu)}=\int_{0}^{\upsilon}{dw}\bigg{[}a_{s}{\widehat{P}}^{(0)}_{gg}{(}\upsilon-w{)}+a^{2}_{s}\widehat{P}^{(1)}_{gg}{(}\upsilon-w{)}\bigg{]}
 \widehat{D}^{T_{nQ}}_{g}(w,\mu),
\end{eqnarray}
where
$\widehat{P}_{gg}(\upsilon){\equiv}e^{-\upsilon}{P}_{gg}(e^{-\upsilon})$.
We define the Laplace transform of a function
$\widehat{P}_{gg}(\upsilon)$ as $\Phi_{g}(s)$, where
\begin{eqnarray}
\Phi_{g}(s){\equiv}\mathcal{L}[{P}_{gg}(\upsilon);
s]=\int_{0}^{\infty}{P}_{gg}(\upsilon)e^{-s\upsilon}d\upsilon,
\end{eqnarray}
and the Laplace transform of a convolution of functions is defined
as
\begin{eqnarray}
\mathcal{L}\bigg{[}\int_{0}^{\upsilon}\widehat{D}^{T_{nQ}}_{g}(w)\hat{P}_{gg}(\upsilon-w)dw;
s\bigg{]}=\Phi_{g}(s){\times}\mathcal{D}_{g}^{T_{nQ}}(s),
\end{eqnarray}
where
$\mathcal{D}_{g}^{T_{nQ}}(s,\mu)=\mathcal{L}[\hat{D}_{g}^{T_{nQ}}(\upsilon,\mu);
s]$. By utilizing the property  that the Laplace transform of a
convolution is equal to the product of the Laplace transforms of
the individual factors, we can simplify Eq.(8) in the $s$-space
\begin{eqnarray}
\dfrac{{\partial}\mathcal{D}^{T_{nQ}}_{g}(s,\mu)}{{\partial}\ln(\mu)}=[a_{s}\Phi_{g}^{(0)}(s)+a^{2}_{s}\Phi_{g}^{(1)}(s)]
 \mathcal{D}^{T_{nQ}}_{g}(s,\mu),
\end{eqnarray}
where the function $\Phi_{g}^{(1)}(s)$ is defined in the
$s$-space in Appendix C.\\

$\bullet$ LO analysis:\\
In the LO analysis, Eq.(11) can be rewritten as follows
\begin{eqnarray}
\dfrac{{\partial}\mathcal{D}^{T_{nQ}}_{g}(s,\mu)}{{\partial}\ln(\mu)}=a_{s}\Phi_{g}^{(0)}(s)
 \mathcal{D}^{T_{nQ}}_{g}(s,\mu),
\end{eqnarray}
where the solution to the equation above is as follows
\cite{Block5}
\begin{eqnarray}
\mathcal{D}^{T_{nQ}}_{g}(s,\tau)=k^{(0)}_{gg}(s,\tau)\mathcal{D}^{T_{nQ}}_{0g}(s),
\end{eqnarray}
where
\begin{eqnarray}
k^{(0)}_{gg}(s,\tau)&{\equiv}&
e^{\frac{\tau}{2}\Phi^{(0)}_{g}(s)}\bigg{[}
\cosh(\frac{\tau}{2}\Phi^{(0)}_{g}(s))+\sinh(\frac{\tau}{2}\Phi^{(0)}_{g}(s))\bigg{]},
\end{eqnarray}
with
\begin{equation}
\Phi_{g}^{(0)}(s)=12(\frac{1}{s}-\frac{2}{s+1}+\frac{1}{s+2}-\frac{1}{s+3}-\psi(s+1)-\gamma_{E})+\frac{33-2n_{f}}{3},
\end{equation}
where $\psi$ is the digamma function and $\gamma_{E}=0.5772156...$
is Euler$^{,}$s constant, and
\begin{eqnarray}
\tau(\mu,\mu_{0})=\int_{q=\mu_{0}}^{q=\mu}\frac{\alpha_{s}(q^{2})}{4\pi}d{\ln}q^{2}.
\end{eqnarray}
By applying the inverse Laplace transform of a product to the
convolution of functions, we have
\begin{eqnarray}
\mathcal{L}^{-1}[\mathcal{D}^{T_{nQ}}_{g}(s,\tau);
\upsilon]&=&\widehat{\mathcal{D}}^{T_{nQ}}_{g}(\upsilon,\tau)\nonumber\\
&&=\mathcal{L}^{-1}[k^{(0)}_{gg}(s,\tau){\times}\mathcal{D}^{T_{nQ}}_{0g}(s);
\upsilon]=\int_{0}^{\infty} \widehat{K}^{(0)}_{GG}(\upsilon-w,
\tau)
 \widehat{\mathcal{D}}^{T_{nQ}}_{0g}(w)dw,
\end{eqnarray}
where
$\widehat{K}^{(0)}_{GG}(\upsilon,\tau){\equiv}\mathcal{L}^{-1}[k_{gg}^{(0)}(s,\tau);
\upsilon]$ and $\widehat{D}^{T_{nQ}}_{0g}(\upsilon)$ represents
the FF of a gluon at the initial scale. Transforming the equation
above back into the $x$ space, we find the FF at the scale $\mu$,
into the initial scale $\mu_{0}$ and the lowest-order
contributions to the FFs (please refer to Appendix A), by
\begin{eqnarray}
D^{T_{nQ}}_{g}(z,\mu)=\int_{z}^{1}{\frac{d\xi}{\xi}}K_{GG}^{(0)}\bigg{(}{\ln}(\frac{z}{\xi}),\tau(\mu,\mu_{0})\bigg{)}
 D^{T_{nQ}}_{0g}(\xi),
\end{eqnarray}

$\bullet$ NLO analysis:\\
In the NLO analysis, the dependence of the NLO terms on the
right-hand side of Eq.(11) on $\tau$ preserves a second Laplace
transform into $U$-space \cite{Block6}. This can be represented as
where
\begin{eqnarray}
\mathcal{G}^{T_{nQ}}_{g}(s,U)&=&\mathcal{L}[\mathcal{D}^{T_{nQ}}_{g}(s,\tau);
U],\nonumber\\
\mathcal{L}[\frac{{\partial}\mathcal{D}^{T_{nQ}}_{g}}{\partial{\tau}}(s,\tau);
U]&=& U\mathcal{G}^{T_{nQ}}_{g}(s,U)-\mathcal{D}^{T_{nQ}}_{0g}(s).
\end{eqnarray}
In $U$-space, where $s$ is a parameter in $U$-space, the equation
becomes:
\begin{eqnarray}
U\mathcal{G}^{T_{nQ}}_{g}(s,U)-\mathcal{D}^{T_{nQ}}_{0g}(s)=\Phi_{g}^{(0)}(s)\mathcal{G}^{T_{nQ}}_{g}(s,U)
+\Phi_{g}^{(1)}(s)\mathcal{L}[a_{s}(\tau)\mathcal{D}^{T_{nQ}}_{g}(s,\tau);U].
\end{eqnarray}
Here, $a_{s}(\tau)=\frac{\alpha_{s}(\tau)}{4\pi}$ with an
approximation given by $
a_{s}(\tau){\approx}~a_{0}+a_{1}e^{-b_{1}\tau}$, where the
constants are detailed in Ref.[35] as $a_{0}=0.0037$,
$a_{1}=0.025$ and $b_{1}=10.7$. This equation can be rewritten in
the following form
\begin{eqnarray}
[U-\Phi_{g}^{(0)}(s)-a_{0}\Phi_{g}^{(1)}(s)]\mathcal{G}^{T_{nQ}}_{g}(s,U)=\mathcal{D}^{T_{nQ}}_{0g}(s)
+a_{1}\Phi_{g}^{(1)}(s)\mathcal{G}^{T_{nQ}}_{g}(s,U+b_{1}),
\end{eqnarray}
where $\mathcal{L}[a_{s}(\tau)\mathcal{D}^{T_{nQ}}_{g}(s,\tau);U]
=a_{0}\mathcal{G}^{T_{nQ}}_{g}(s,U)+a_{1}\mathcal{G}^{T_{nQ}}_{g}(s,U+b_{1})
$. The solution to Eq.(21) is obtained in the following manner
\begin{eqnarray}
\mathcal{G}^{T_{nQ}}_{g}(s,U)=\frac{1}{[U-\Phi_{g}^{(0)}(s)-a_{0}\Phi_{g}^{(1)}(s)]}\bigg{[}\mathcal{D}^{T_{nQ}}_{0g}(s)
+a_{1}\Phi_{g}^{(1)}(s)\mathcal{G}^{T_{nQ}}_{g}(s,U+b_{1})\bigg{]}.
\end{eqnarray}
Because the terms in $\mathcal{G}^{T_{nQ}}_{g}(s,U)$ decrease
rapidly as $1/|U|$ for $|U|{\rightarrow}\infty$, we can now apply
the Laplace inverse from $U$ space to $\tau$ space, as
\begin{eqnarray}
\mathcal{D}^{T_{nQ}}_{g}(s,\tau)=k^{(1)}_{gg}(a_{1},b_{1},s,\tau)\mathcal{D}^{T_{nQ}}_{0g}(s).
\end{eqnarray}
The function $k^{(1)}_{gg}(a_{1}, b_{1},s,\tau)$ can be expressed
as a power series in the NLO expansion parameter $a_{1}$. The
coefficients of this series are analytic functions of $s$ and
$\tau$, and can be represented by the following form \cite{R17}:
\begin{eqnarray}
k^{(1)}_{gg}(a_{1},
b_{1},s,\tau)&=&\{e^{-\frac{1}{2}b_{1}\tau}[b_{1}((3\Phi_{g}-b_{1}^{2})e^{\tau(b_{1}+\Phi_{g})})+4a_{1}e^{\frac{1}{2}\tau(b_{1}+\Phi_{g})}(\Phi_{g}^{4}\cosh(\frac{1}{2}\tau{\Phi_{g}})
\sinh(\frac{1}{2}\tau{b_{1}})\nonumber\\
&&-\Phi^{2}_{g}(b_{1}^{2}-\Phi^{2}_{g})\sinh(\frac{1}{2}\tau{\Phi_{g}}))]\}/(2b_{1}\Phi_{g}(\Phi^{2}_{g}-b_{1}^{2})),
\end{eqnarray}
with $\Phi_{g}$ is defined as $\Phi_{g}{\equiv}\Phi^{(0)}_{g}(s)+
a_{0}\Phi^{(1)}_{g}(s)$. After applying the inverse Laplace
transform for  $s$ to $\upsilon$ spaces and transforming back into
$x$ space, we find that:
\begin{eqnarray}
D^{T_{nQ}}_{g}(z,\mu)=\int_{z}^{1}{\frac{d\xi}{\xi}}K^{(1)}_{GG}\bigg{(}{\ln}(\frac{z}{\xi}),\tau(\mu,\mu_{0})\bigg{)}
 D^{T_{nQ}}_{0g}(\xi),
\end{eqnarray}
where
$K^{(1)}_{GG}(\upsilon,\tau){\equiv}\mathcal{L}^{-1}[k^{(1)}_{gg}(a_{1},b_{1},s,\tau);
\upsilon]$. So, we have an explicit solution for the FF of a gluon
at the LO and NLO  approximations, which can be evaluated with the
numerical accuracy of $D^{T_{nQ}}_{0g}(z)$  known for
$[g{\rightarrow}T_{2Q}]$ and $[g{\rightarrow}T_{4Q}]$. With an
analytical FF of a gluon at the initial scale, one can numerically
extract the FF of a gluon at any desired $z$ and $\mu^2$
values.\\

\subsection{III. Results and Discussions}

With the explicit form of gluon fragmentation in the channel
leading to the $g{\rightarrow}T_{nQ}$ at the initial scale inputs,
we can now extract numerical results for $\mu>\mu_{0}$. This is
done using the FF of a gluon at  initial scales of
$\mu_{0}=2m_{c}$, $4m_{c}$ and $4m_{b}$ as reported in
Refs.\cite{R1} and \cite{R22}. The QCD parameter $\Lambda$ for
active flavors has been determined \cite{R28} based on
$\alpha_{s}(M_{z}^{2})=0.1166$ as
\begin{eqnarray}
\mathrm{LO} :
\Lambda(n_{f}=3)=0.1368~\mathrm{GeV},~\Lambda(n_{f}=4)=0.1368~\mathrm{GeV},
\Lambda(n_{f}=5)=0.8080~\mathrm{GeV},\nonumber\\
\mathrm{NLO} :
\Lambda(n_{f}=3)=0.3472~\mathrm{GeV},~\Lambda(n_{f}=4)=0.2840~\mathrm{GeV},
\Lambda(n_{f}=5)=0.1957~\mathrm{GeV}.
\end{eqnarray}
\begin{figure}
\centerline{
\includegraphics[width=0.5\textwidth]{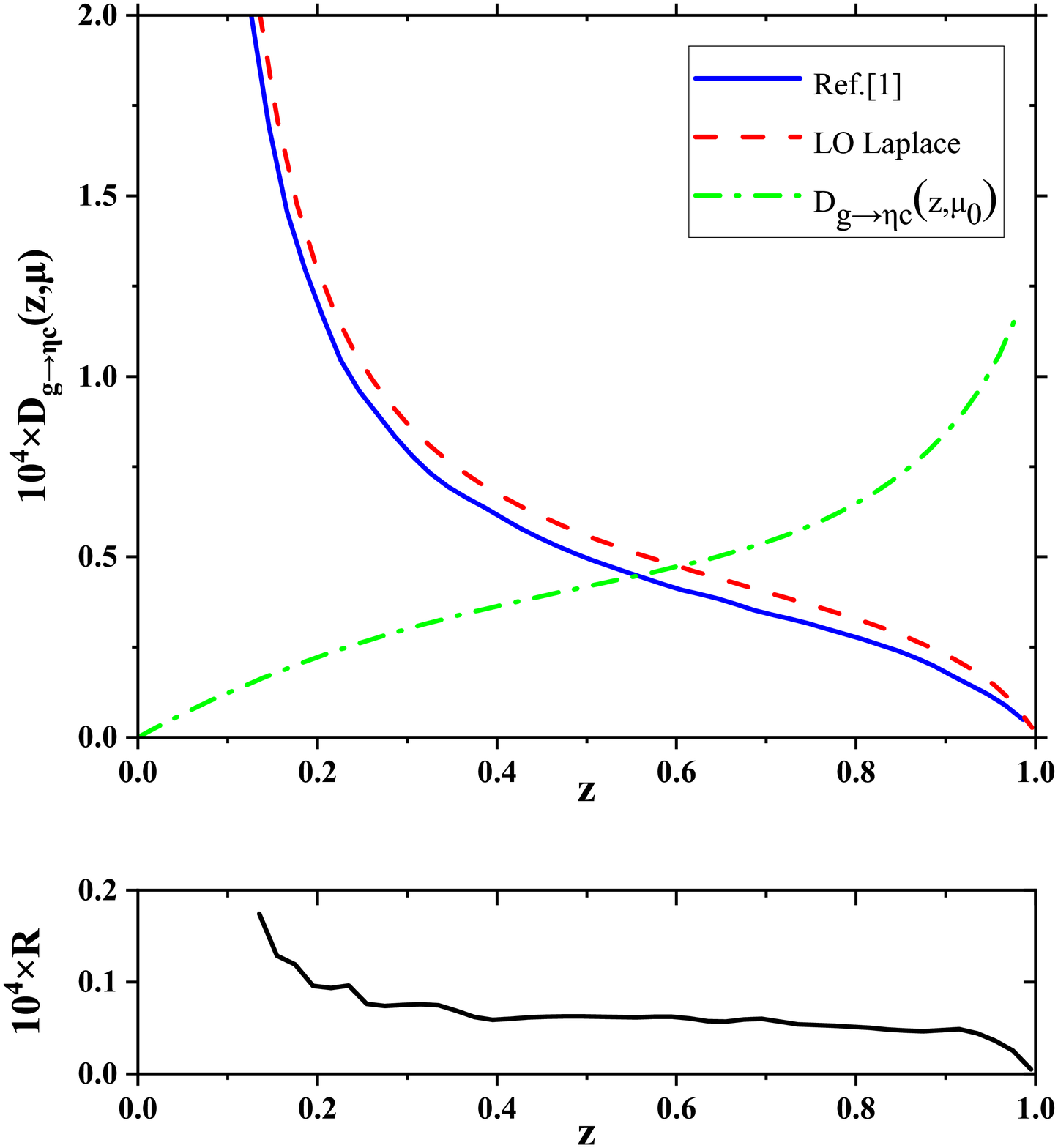}}
\caption{Evolution of the fragmentation function
$D_{g{\rightarrow}\eta_{c}}(z,\mu)$ as a function of the variable
$z$ is shown through Laplace transform for $\mu=20m_{c}$ at the LO
approximation (red-dashed curve), compared with the FF in
Ref.\cite{R1} (blue-solid curve). The green-dashed-dot curve
represents the FF of a gluon at the initial scale $\mu=2m_{c}$
according to Ref.\cite{R1}. The difference
$R=D_{g{\rightarrow}\eta_{c}}(\mathrm{Laplace}-\mathrm{Bratten})(z,20m_{c})$
between the results in the LO approximation and the Braaten
 results \cite{R1} is shown in the lower panel over a wide range of
$z$. }\label{Fig1}
\end{figure}
\begin{figure}
\centerline{
\includegraphics[width=0.5\textwidth]{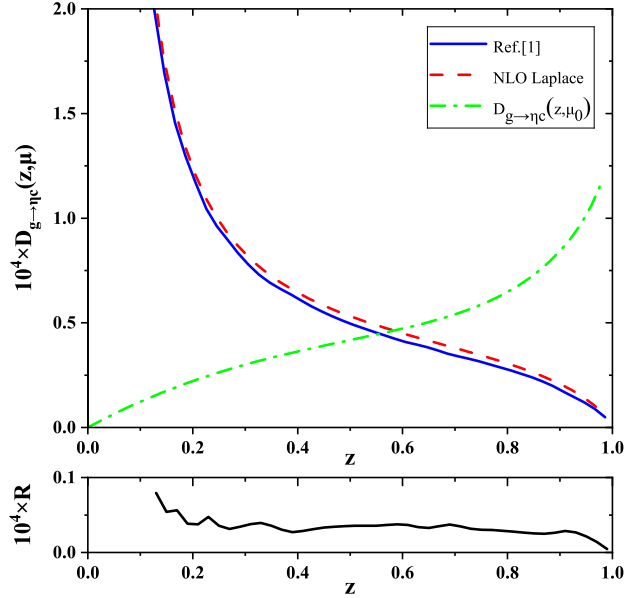}}
\caption{ The same as Fig.1 at the NLO approximation.}\label{Fig2}
\end{figure}
In Fig.1, the results for the FF of a gluon in the channel
$g{\rightarrow}T_{2c}$ within the LO approximation are shown and
compared with the results of Ref.\cite{R1} at $\mu=20m_{c}$. To
demonstrate the effects of evolution, we evolve the FF at the
scale $\mu=20m_{c}$ with respect to the initial FF
$D_{g{\rightarrow}\eta_{c}}(z,\mu=2m_{c})$. The $z$ dependence of
$D_{g{\rightarrow}\eta_{c}}(z,\mu)$ at the initial scale
$\mu=2m_{c}$ and the evolved scale  $\mu=20m_{c}$ at the LO
approximation is shown in Fig.1. Additionally, in Fig.2, the
results for the FF of a gluon in the channel
$g{\rightarrow}T_{2c}$ within the NLO approximation are shown and
compared with the results of Ref.\cite{R1} at $\mu=20m_{c}$ as
shown in Fig.1. The difference between the results in the LO and
NLO approximations with the Braaten \cite{R1} results are shown in
the lower panel of the Figs.1 and 2, respectively. As observed,
the gluon fragmentation at NLO accuracy using the Laplace
transform is closer to the result in Ref.\cite{R1} than the
calculations at LO in a wide range of $z$. These differences
indicate that our results are in good agreement with other models
in the low-$z$ region. In the following, we apply the NLO
corrections to the calculations of the gluon
fragmentation functions.\\
In Fig.3, the evolution of the fragmentation function
$D_{g{\rightarrow}\eta_{c}}(z,\mu)$ by Laplace transform at the
NLO approximation is shown over a wide range of $\mu=10-80m_{c}$.
We observe that the point $z{\simeq}0.4$ is independent of the
evolution. The fragmentation function
$D_{g{\rightarrow}\eta_{c}}(z,\mu)$ increases as $\mu$ increases
and $z$ decreases (i.e., $z{\lesssim}0.4$). It decreases as $\mu$
and $z$
increase (i.e., $z{\gtrsim}0.4$).\\
\begin{figure}
\centerline{
\includegraphics[width=0.5\textwidth]{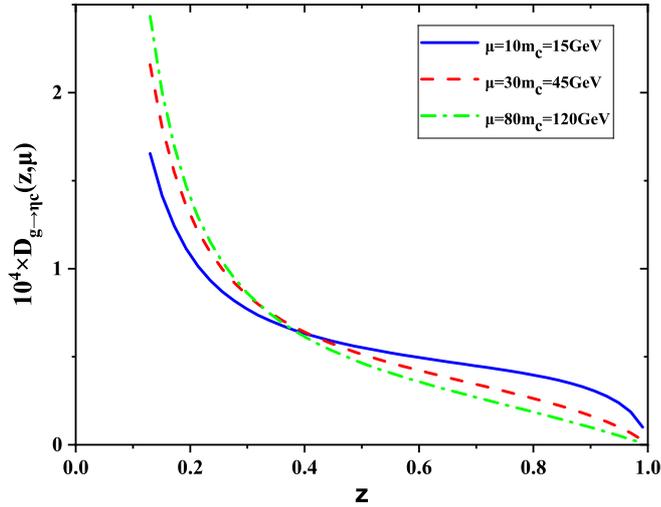}}
\caption{The fragmentation functions
$D_{g{\rightarrow}\eta_{c}}(z,\mu)$ are plotted at fixed $\mu$
values: $\mu=10m_{c}$ (solid-blue curve), $\mu=30m_{c}$
(dashed-red curve), and $\mu=80m_{c}$ (dashed dot-green curve).
The plot shows the FF as a function of the variable $z$ at the NLO
approximation.}\label{Fig3}
\end{figure}
In Figs.4 and 5, the gluon fragmentation channels to the
$T_{4c}(0^{++})$ and $T_{4c}(2^{++})$ states as a function of the
variable $z$ using the Laplace transform method are presented at
the NLO approximation and compared with those in Ref.\cite{R22} at
the scale $\mu=5m_{c}$. The results are evaluated based on the
gluon fragmentation at the initial scale $\mu=4m_{c}$ as defined
in Eqs.(27) and (28) in Appendix A respectively. The difference
between the results in the NLO approximation for the gluon
fragmentation channels to the $T_{4c}(0^{++})$ and
$T_{4c}(2^{++})$ states compared to the results by F.G. Celiberto
et al., \cite{R22} are shown in the
lower panel of Figs.4 and 5, respectively.\\
\begin{figure}
\centerline{
\includegraphics[width=0.5\textwidth]{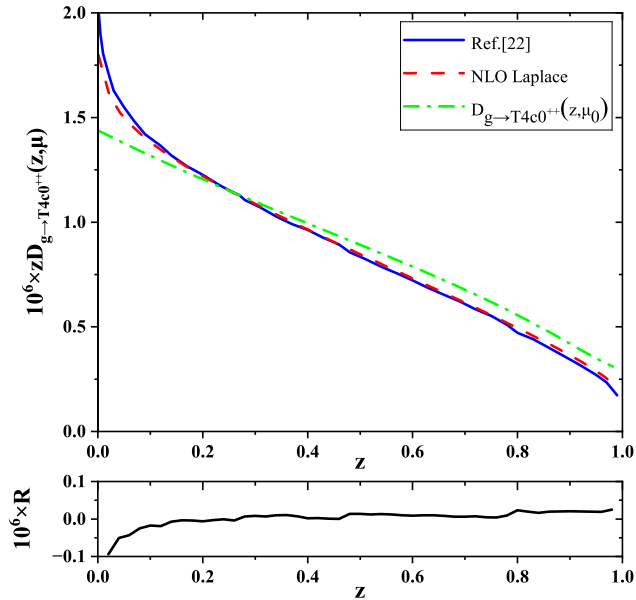}}
\caption{Evolution of the fragmentation function
$D_{g{\rightarrow}T_{4c}(0^{++})}(z,\mu)$ for $T_{4c}(0^{++})$
state is shown as a function of the variable $z$ using Laplace
transform for $\mu=5m_{c}$ at the NLO approximation (red-dashed
curve), compared with the FF in Ref.\cite{R22} (blue-solid curve).
The green-dashed-dot curve represents the FF of a gluon at the
initial scale $\mu=4m_{c}$ according to Ref.\cite{R22} for the
$T_{4c}(0^{++})$ state. The difference
$R=D_{g{\rightarrow}T_{4c}(0^{++})}(\mathrm{Laplace}-\mathrm{Celiberto})(z,5m_{c})$
between the results in the NLO approximation and the results from
F.G. Celiberto et al., \cite{R22} is shown in the lower panel over
a wide range of $z$.}\label{Fig4}
\end{figure}
\begin{figure}
\centerline{
\includegraphics[width=0.5\textwidth]{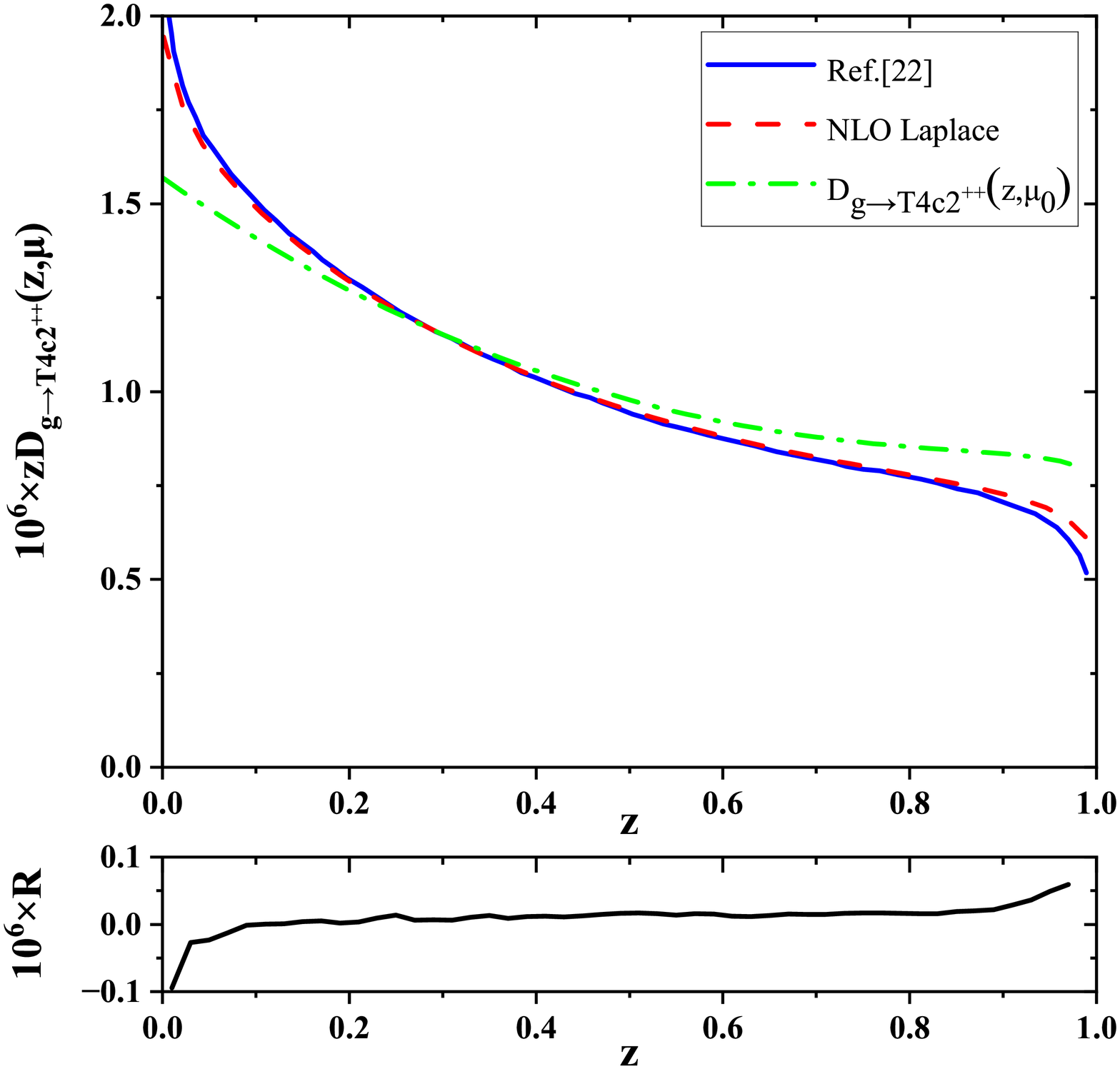}}
\caption{ The same as Fig.4 for the gluon fragmentation channel to
the $T_{4c}(2^{++})$ state. }\label{Fig5}
\end{figure}
In Figs.6 and 7, the gluon fragmentation channels to the
$T_{4b}(0^{++})$ and $T_{4b}(2^{++})$ states as a function of the
variable $z$ by Laplace transform are presented at the NLO
approximation and compared with those in Ref.\cite{R23} at the
scale $\mu=5m_{c}$. The results are evaluated according to the
gluon fragmentation at the initial scale $\mu=4m_{b}$ as defined
in Eqs.(35) and (36) in Appendix A,respectively.\\
\begin{figure}
\centerline{
\includegraphics[width=0.5\textwidth]{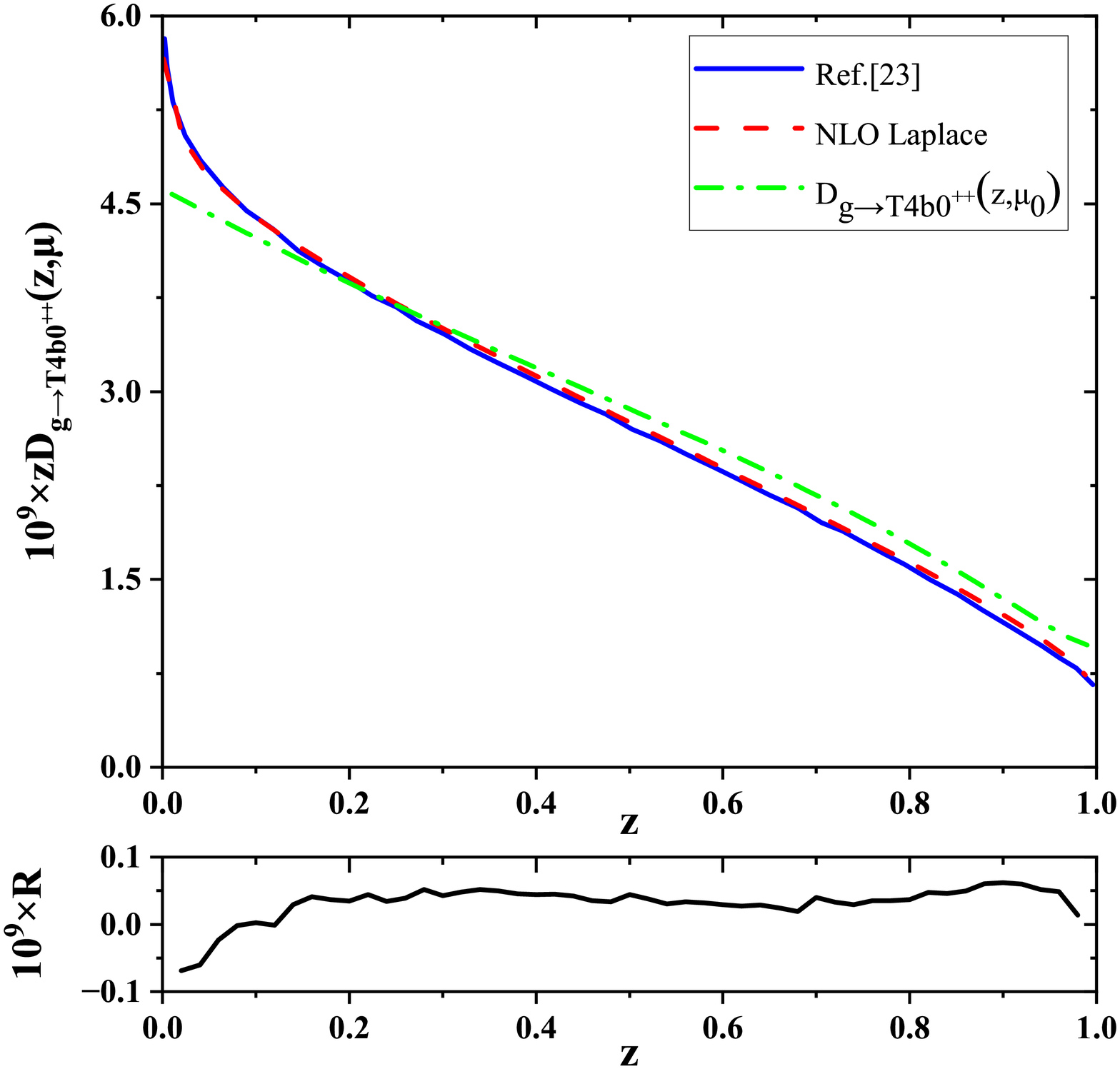}}
\caption{Evolution of the fragmentation function
$D_{g{\rightarrow}T_{4b}(0^{++})}(z,\mu)$ for $T_{4b}(0^{++})$
state is shown as a function of the variable $z$ using Laplace
transform for $\mu=5m_{b}$ at the NLO approximation (red-dashed
curve), compared with the FF in Ref.\cite{R23} (blue-solid curve).
The green-dashed-dot curve represents the FF of a gluon at the
initial scale $\mu=4m_{b}$ according to Ref.\cite{R23} for the
$T_{4b}(0^{++})$ state. The difference
$R=D_{g{\rightarrow}T_{4b}(0^{++})}(\mathrm{Laplace}-\mathrm{Celiberto})(z,5m_{b})$
between the results in the NLO approximation and the results from
F.G. Celiberto et al., \cite{R23} is shown in the lower panel over
a wide range of $z$.}\label{Fig6}
\end{figure}
\begin{figure}
\centerline{
\includegraphics[width=0.5\textwidth]{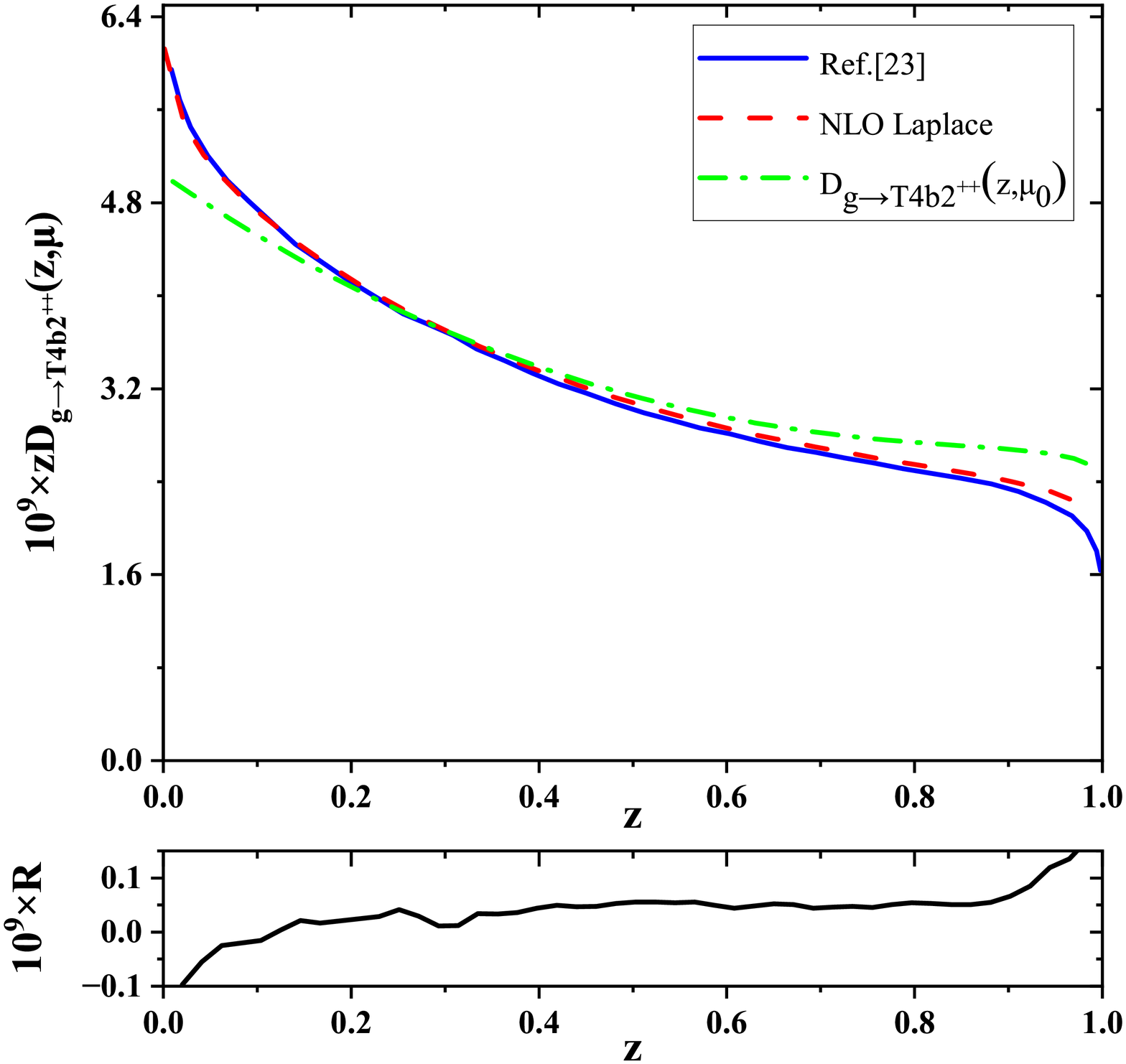}}
\caption{ The same as Fig.6 for the gluon fragmentation channel to
the $T_{4b}(2^{++})$ state.}\label{Fig7}
\end{figure}
In Figs.4-7, we present a comparison between the FF of a gluon,
$g{\rightarrow}T_{4Q}$ in Laplace transform for the ground state
($0^{++}$) and the radial excitation ($2^{++}$),which shows
correct behavior compared to the results in Refs.\cite{R22, R23}.
For clarity, in Fig.8 we display the $z$-dependence of our
$D_{g}^{T_{4c}(0^{++})}$ using the Laplace evolution method at
various scales $\mu=10, 20$, and $50m_{c}$, which can also be
applied to the radial excitations $T_{4c}(2^{++})$ for fully
charmed tetraquarks and fully bottomed tetraquarks. The evolution
of the fragmentation function $D_{g{\rightarrow}\eta_{c}}(z,\mu)$
through Laplace transform at the NLO approximation is illustrated
in Fig.8 across a wide range of $\mu=10, 20$ and $ 50m_{c}$. We
notice that the point $z{\simeq}0.2$ remains constant throughout
the evolution. In comparison to Fig.3, this point decreases as the
fragmentation process of the gluon transitions from
$[g{\rightarrow}T_{2Q}]$ to $[g{\rightarrow}T_{4Q}]$. The
fragmentation function $D_{g{\rightarrow}\eta_{c}}(z,\mu)$
increases as $\mu$ increases and $z$ decreases (i.e.,
$z{\lesssim}0.2$) and decreases as both $\mu$ and $z$
increase (i.e., $z{\gtrsim}0.2$).\\
\begin{figure}
\centerline{
\includegraphics[width=0.5\textwidth]{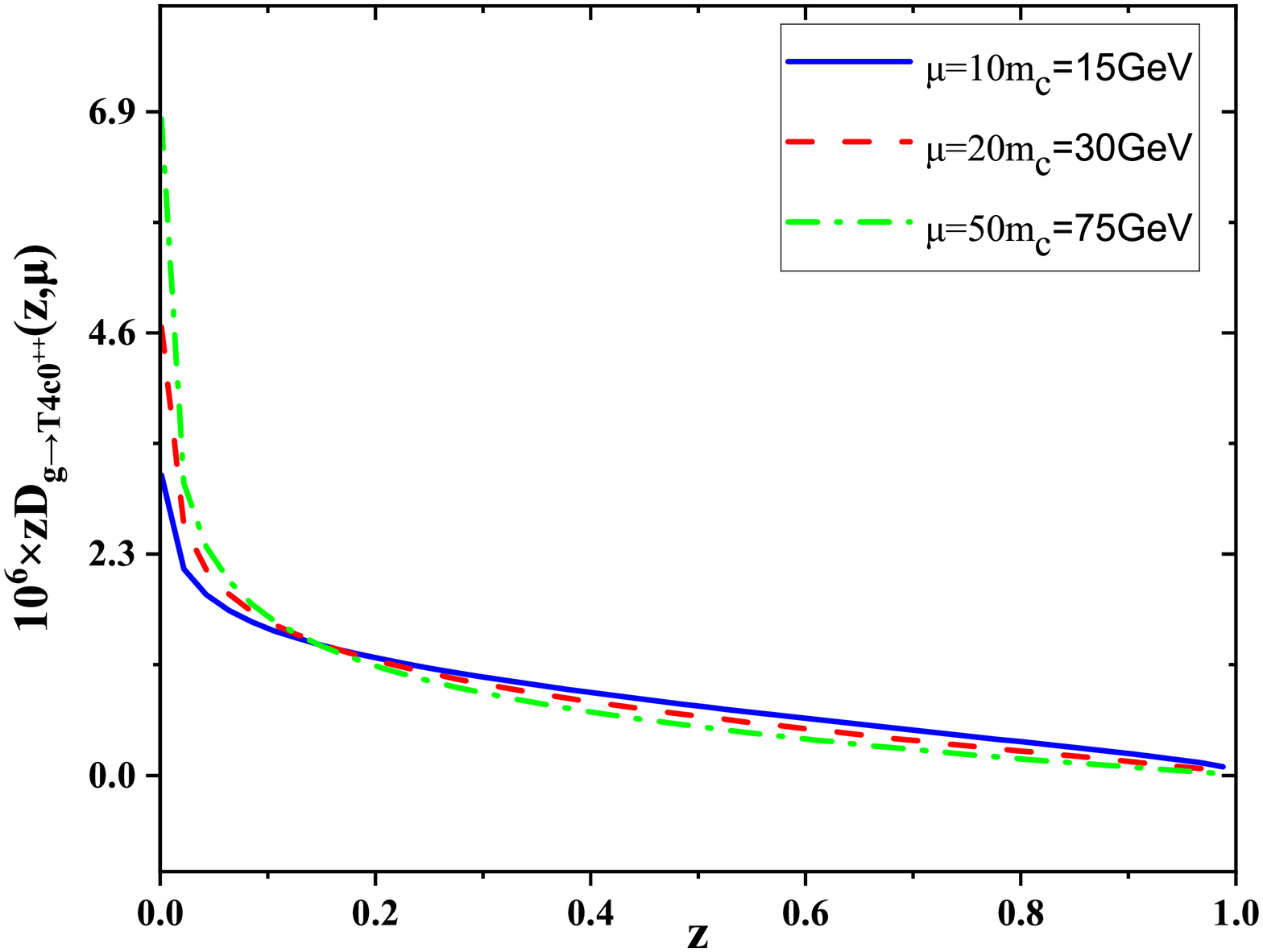}}
\caption{The evolution of the gluon fragmentation channel to the
$T_{4c}(0^{++})$ state as a function of the variable $z$ for
$\mu=10m_{c}$(blue-solid curve), $\mu=20m_{c}$(red-dashed curve)
and $\mu=50m_{c}$(green-dashed-dot curve) is presented at the NLO
approximation.}\label{Fig8}
\end{figure}

\subsection{IV. Conclusion}

We have introduced a method based on the Laplace transform
technique to analyze the evolution of the fragmentation mechanisms
 $[g{\rightarrow}T_{2Q}]$ and
$[g{\rightarrow}T_{4Q}]$ at both the LO and NLO approximations.
The gluon fragmentation function in the $[g{\rightarrow}T_{2c}]$
channel is calculated for the fragmentation of a gluon splitting
into an $S$-wave charmonium state starting from $\mu=2m_{c}$. The
evolution of the fragmentation function for the semi-inclusive
detection of fully charmed and bottomed tetraquarks is then
examined, including bound states of $T_{4Q}(0^{++})$ and their
$T_{4Q}(2^{++})$ radial excitations. This analysis is based on
single-parton collinear fragmentation in a VFNS and relies on the
parametrization of the fragmentation functions within a
kinematical region characterized by moderate and large transverse
momenta at the initial scales. We find that the Laplace transform
method, used in the evolution of the fragmentation function at the
NLO approximation, provides correct behaviors across a wide range
of $z$ and $\mu$
scales.\\

\subsection{Appendix A}

The lowest-order contributions to the FF of a gluon describe the
perturbative part of the $[g{\rightarrow}T_{4c}(0^{++})]$ and
$[g{\rightarrow}T_{4c}(2^{++})]$. The composite quantum number
$[n]$, which runs over the combinations [3, 3], [6, 6], [3, 6] and
[6, 3], is defined by the following forms, respectively
\cite{R22}:
\begin{eqnarray}
D_{g}^{T_{4c}(0^{++})}(z,4m_{c})=\frac{\mathrm{GeV}^{9}}{m_{c}^{9}}\bigg{[}0.0347\mathcal{D}_{g}^{(0^{++})}
(z,[3,3])+0.0128\mathcal{D}g^{(0^{++})} (z,[6,6])+
0.0422\mathcal{D}_{g}^{(0^{++})} (z,[3,6]) \bigg{]},
\end{eqnarray}
and
\begin{eqnarray}
D_{g}^{T_{4c}(2^{++})}(z,4m_{c})=\frac{\mathrm{GeV}^{9}}{m_{c}^{9}}\bigg{[}0.072\mathcal{D}_{g}^{(2^{++})}
(z,[3,3]) \bigg{]},
\end{eqnarray}
with $m_{c}=1.5~\mathrm{GeV}$ being the mass of the charm quark.
The SDSs read
\begin{eqnarray}
D_{g}^{(0^{++})}(z,[3,3])&=&\frac{\pi^2\alpha_{s}^{4}(4m_{c})}{497664z(2-z)^{2}(3-z)}
[186624-430272z+511072z^2-425814z^3+217337z^4-61915z^5+7466z^6\nonumber\\
&&-42(1-z)(2-z)(3-z)(-144+634z-358z^2+70z^3)\ln(1-z)
+36(2-z)(3-z)(144-634z\nonumber\\
&&+749z^2-364z^3+74z^4)\ln(1-\frac{z}{2})+12(2-z)(3-z)(72-362z+361z^2-136z^3+23z^4)\ln(1-\frac{z}{3})],\nonumber\\
\end{eqnarray}

\begin{eqnarray}
D_{g}^{(0^{++})}(z,[6,6])&=&\frac{\pi^2\alpha_{s}^{4}(4m_{c})}{331776z(2-z)^{2}(3-z)}[186624-430272z+617824z^2-634902z^3\nonumber\\
&&+374489z^4-115387z^5+14387z^6-6(1-z)(2-z)(3-z)(-144-2166z+1015z^2+70z^3)\ln(1-z)\nonumber\\
&&-156(2-z)(3-z)(144-1242z+1693z^2-876z^3+170z^4)\ln(1-\frac{z}{2})+300(2-z)(3-z)(72\nonumber\\
&&-714z+953z^2-472z^3+87z^4)\ln(1-\frac{z}{3})],
\end{eqnarray}

\begin{eqnarray}
D_{g}^{(0^{++})}(z,[3,6])&=&\frac{\pi^2\alpha_{s}^{4}(4m_{c})}{165888z(2-z)^{2}(3-z)}[186624-430272z+490720z^2-394422z^3+199529z^4-57447z^5+7082z^6\nonumber\\
&&+6(1-z)(2-z)(3-z)(-432+3302z-1855z^2+210z^3)\ln(1-z)-12(2-z)(3-z)(720\nonumber\\
&&-2258z+2329z^2-1052z^3+226z^4)\ln(1-\frac{z}{2})+12(2-z)(3-z)(936-4882z\nonumber\\
&&+4989z^2-1936z^3+331z^4)\ln(1-\frac{z}{3})],
\end{eqnarray}
\begin{eqnarray}
D_{g}^{(2^{++})}(z,[3,3])&=&\frac{\pi^2\alpha_{s}^{4}(4m_{c})}{622080z^2(2-z)^{2}(3-z)}[(46656-490536z+1162552z^2-1156308z^3+595421z^4-170578z^5\nonumber\\
&&+21212z^6)2z+3(1-z)(2-z)(3-z)(-20304-31788z)(1296+1044z+73036z^2-36574z^3\nonumber\\
&&+7975z^4)\ln(1-z)+33(2-z)(3-z)(1296+25)(-9224z^2+9598z^3
-3943z^4+725z^5)\ln(1-\frac{z}{3})],\nonumber\\
\end{eqnarray}
\begin{eqnarray}
D_{g}^{(2^{++})}(z,[6,6])=0,
\end{eqnarray}
 \begin{eqnarray}
D_{g}^{(2^{++})}(z,[3,6])=0.
\end{eqnarray}

The lowest-order contributions to the FF of a gluon describe the
perturbative part of the $[g{\rightarrow}T_{4b}(0^{++})]$ and
$[g{\rightarrow}T_{4b}(2^{++})]$ processes. They are defined by
the following forms, respectively \cite{R22}:
\begin{eqnarray}
D_{g}^{T_{4b}(0^{++})}(z,4m_{b})=\frac{\mathrm{GeV}^{9}}{m_{b}^{9}}\bigg{[}13.88\mathcal{D}_{g}^{(0^{++})}
(z,[3,3])+5.12\mathcal{D}g^{(0^{++})} (z,[6,6])+
16.88\mathcal{D}_{g}^{(0^{++})} (z,[3,6]) \bigg{]},
\end{eqnarray}
and
\begin{eqnarray}
D_{g}^{T_{4b}(2^{++})}(z,4m_{b})=\frac{\mathrm{GeV}^{9}}{m_{b}^{9}}\bigg{[}28.8\mathcal{D}_{g}^{(2^{++})}
(z,[3,3]) \bigg{]},
\end{eqnarray}
with $m_{b}=4.9~\mathrm{GeV}$ being the mass of the bottom
quark.\\

\subsection{Appendix B}

The splitting functions at the LO and NLO approximations read
\begin{equation}
P_{gg}^{(0)}(z)=12[\frac{z}{(1-z)}_{+}+\frac{1-z}{z}+z(1-z)]+\frac{33-2n_{f}}{3}\delta(1-z)
\end{equation}
 where the plus function is defined as
\begin{equation}
\int_{x}^{1}\frac{f(z)}{(1-z)}_{+}dz=\int_{x}^{1}\frac{f(z)-f(1)}{1-z}dz+f(1)\ln(1-x)
\end{equation}

\begin{eqnarray}
P_{gg}^{(1)}(z)&=&2C_{F}T_{f}\{-4+12z-\frac{164}{9}z^{2}+(10+14z+\frac{16}{3}z^{2}+\frac{16}{3z})\ln(z)+\frac{92}{9z}+2(1+z)\ln^{2}(z)\}\nonumber\\
&&+2C_{A}T_{f}\{2-2z+\frac{26}{9}(z^{2}-\frac{1}{z})-\frac{4}{3}(1+z)\ln(z)-(\frac{20}{9}+\frac{8}{3}{\ln}z)P_{gg}(z)\}\nonumber\\
&&+2C_{A}^{2}\{\frac{27}{2}(1-z)+\frac{67}{9}(z^{2}-\frac{1}{z})+(\frac{11}{3}-\frac{25}{3}z-\frac{44}{3z})\ln(z)-4(1+z)\ln^{2}(z)+2P_{gg}(-z)S_{2}(z)\nonumber\\
&&+[4\ln(z)\ln(1-z)-3\ln^{2}(z)+\frac{22}{3}\ln(z)-\frac{\pi^{2}}{3}+\frac{67}{9}]P_{gg}(z)\},
\end{eqnarray}
where for the specific case of SU(3) we have
\begin{eqnarray}
C_{A}=3, C_{F}=4/3, T_{R}=1/2.
\end{eqnarray}
Here $P_{gg}(z)=1/(1-z)+1/z+z(1-z)-2$ and
$S_{2}(z)=\int_{z/(1+z)}^{1/(1+z)}dy\ln[(1-y)/y]/y=-2\mathrm{Li}_{2}(-z)+(1/2)\ln^{2}(z)-2\ln(z)\ln(1+z)-\pi^{2}/6$.\\
%%%%%%%%%%%%%%%%%%%%%%%%%%%%%%%%%%%%%%%%%%%%%%

\subsection{Appendix C}

The Laplace transform of the gluon-gluon splitting function,
denoted by $\Phi_g^{(1)}$ at the NLO approximation reads
\cite{R17}:
\begin{eqnarray}
\Phi_g^{(1)}(s)&=&C_{F}T_f\{-\frac{16}{3s^2}+\frac{92}{9s}+\frac{4}{(1+s)^3}-\frac{10}{(1+s)^2}-\frac{4}{1+s}+\frac{4}{(2+s)^3}-\frac{14}{(2+s)^2}+\frac{12}{2+s}-\frac{16}{3(3+s)^2}-\frac{164}{9(3+s)}\}\nonumber\\
&&+C_{A}T_f\{\frac{8}{3s^2}-\frac{46}{9s}-\frac{4}{(1+s)^2}+\frac{58}{9(1+s)}+\frac{4}{(2+s)^2}-\frac{38}{9(2+s)}-\frac{8}{3(3+s)^2}+\frac{46}{9(3+s)}+\frac{8}{3}\psi'(1+s)\}\nonumber\\
&&+C_A^2\{-\frac{8}{s^3}+\frac{22}{3s^2}+\frac{2}{(1+s)^3}+\frac{11}{(1+s)^2}+\frac{4.4407}{1+s}-\frac{17.9984}{(2+s)^3}+\frac{1}{(2+s)^2}-\frac{6.9024}{2+s}-\frac{\pi^2}{3(2+s)}+\frac{5.9702}{(3+s)^3}\nonumber\\
&&+\frac{22}{3(3+s)^2}-\frac{6.7917}{3+s}+\frac{\pi^2}{3(3+s)}-\frac{1.801}{(4+s)^3}-\frac{3.5389}{4+s}+\frac{1.3242}{(5+s)^3}+\frac{1.2736}{5+s}-\frac{0.6348}{(6+s)^3}-\frac{5.6479}{6+s}+\frac{0.1398}{(7+s)^3}\nonumber\\
&&+\frac{9.2228}{7+s}-\frac{7.6863}{8+s}+\frac{6.5284}{9+s}-\frac{3.1431}{10+s}+\frac{0.9985}{11+s}-\frac{0.1537}{12+s}-\frac{67}{9}H_{s}+\frac{\pi^2}{3}H_{s}-\frac{1}{s^2}\{\ln(16)-2\psi(1+\frac{s}{2})\nonumber\\
&&+2\psi(\frac{1+s}{2})+s\psi\prime(1+\frac{s}{2})-s\psi\prime(\frac{1+s}{2})\}+2\{\frac{4}{(1+s)^3}-\frac{\ln(4)}{(1+s)^2}-\frac{\psi(1+\frac{s}{2})}{(1+s)^2}+\frac{\psi(\frac{1+s}{2})}{(1+s)^2}\nonumber\\
&&+\frac{\psi\prime(1+\frac{s}{2})}{2(1+s)}-\frac{\psi\prime(\frac{1+s}{2})}{2(1+s)}\}-\frac{1.9992}{(2+s)^3}\{\frac{16}{(1+s)^2}+\frac{12s}{(1+s)^2}+(2+s)\ln(16)-2(2+s)\psi(1+\frac{s}{2})\nonumber\\
&&+2(2+s)\psi(\frac{1+s}{2})+(2+s)^2\psi\prime(1+\frac{s}{2})-(2+s)^2\psi\prime(\frac{1+s}{2})\}+\frac{0.0149}{(1+s)^3}\{\frac{164}{(1+s)^2(2+s)^2}+\frac{284s}{(1+s)^2(2+s)^2}\nonumber\\
&&+\frac{188s^2}{(1+s)^2(2+s)^2}+\frac{60s^3}{(1+s)^2(2+s)^2}+\frac{8s^4}{(1+s)^2(2+s)^2}-4(3+s)\ln(2)-2(3+s)\psi(1+\frac{s}{2})+2(3+s)\psi(\frac{1+s}{2})\nonumber\\
&&+(3+s)^2\psi\prime(1+\frac{s}{2})-(3+s)^2\psi\prime(\frac{1+s}{2})\}-\frac{0.9005}{(4+s)^3}\{\frac{2176}{(1+s)^2(2+s)^2(3+s)^2}+\frac{4392s}{(1+s)^2(2+s)^2(3+s)^2}\nonumber\\
&&+\frac{3504s^2}{(1+s)^2(2+s)^2(3+s)^2}+\frac{1408s^3}{(1+s)^2(2+s)^2(3+s)^2}+\frac{288s^4}{(1+s)^2(2+s)^2(3+s)^2}+\frac{24s^5}{(1+s)^2(2+s)^2(3+s)^2}\nonumber\\
&&+4(4+s)\ln(2)-2(4+s)\psi(1+\frac{s}{2})+2(4+s)\psi(\frac{1+s}{2})+(4+s)^2\psi\prime(1+\frac{s}{2})-(4+s)^2\psi\prime(\frac{1+s}{2})\}\nonumber\\
&&-\frac{0.6621}{(5+s)^3}\{\frac{57328}{(1+s)^2(2+s)^2(3+s)^2(4+s)^2}+\frac{146144s}{(1+s)^2(2+s)^2(3+s)^2(4+s)^2}+\frac{162160s^2}{(1+s)^2(2+s)^2(3+s)^2(4+s)^2}\nonumber\\
&&+\frac{103728s^3}{(1+s)^2(2+s)^2(3+s)^2(4+s)^2}+\frac{42144s^4}{(1+s)^2(2+s)^2(3+s)^2(4+s)^2}+\frac{11160s^5}{(1+s)^2(2+s)^2(3+s)^2(4+s)^2}\nonumber\\
&&+\frac{1880s^6}{(1+s)^2(2+s)^2(3+s)^2(4+s)^2}+\frac{184s^7}{(1+s)^2(2+s)^2(3+s)^2(4+s)^2}+\frac{8s^8}{(1+s)^2(2+s)^2(3+s)^2(4+s)^2}\nonumber\\
&&-4(5+s)\ln(2)-2(5+s)\psi(1+\frac{s}{2})+2(5+s)\psi(\frac{1+s}{2})+(5+s)^2\psi\prime(1+\frac{s}{2})-(5+s)^2\psi\prime(\frac{1+s}{2})\}\nonumber\\
&&+\frac{4}{s^3}(1+s\gamma_{E}+s(\psi(s)-s\psi\prime(1+s)))-\frac{8}{(s+1)^2}(\gamma_{E}+\frac{1}{1+s}+\psi(1+s)-(1+s)\psi\prime(2+s))\nonumber\\
&&+\frac{4}{(s+2)^2}(\gamma_E+\frac{1}{2+s}+\psi(2+s)-(2+s)\psi\prime(3+s))-\frac{4}{(s+3)^2}(\gamma_E+\frac{1}{3+s}+\psi(3+s)
-(3+s)\psi\prime(4+s))\nonumber\\
&&-\frac{0.3174}{(6+s)^2}\{\ln(16)-2\psi(4+\frac{s}{2})+2\psi(\frac{7+s}{2})+(6+s)\psi\prime(4+\frac{s}{2})-(6+s)\psi\prime(\frac{7+s}{2})\}\nonumber\\
&&+\frac{0.0699}{(7+s)^2}\{\ln(16)+2\psi(4+\frac{s}{2})-2\psi(\frac{9+s}{2})-(7+s)\psi\prime(4+\frac{s}{2})+(7+s)\psi\prime(\frac{9+s}{2})\}+4(H_{s}\psi\prime(1+s)\nonumber\\
&&-\frac{1}{2}\psi\prime\prime(1+s))-\frac{22}{3}\psi\prime(1+s)+3\psi\prime\prime(1+s)\},
\end{eqnarray}
where $H_{s}=\psi(s+1)+\gamma_{E}$.\\

%%%%%%%%%%%%%%%%%%%%%%%%%%%%%%%%%%%%%%%%%%%%%%

\subsection{ACKNOWLEDGMENTS}
The authors are grateful to Razi University for the financial
support of this project. G. R. Boroun thanks F. G. Celiberto for
allowing access to the gluon fragmentation channels related to the
initial
scales.\\

%%%%%%%%%%%%%%%%%%%%%%%%%%%%%%%%%%%%%%%%%%%%%%%%%%%

%%%%%%%%%%%%%%%%%%%%%%%%%%%%%%%%%%%%%%%%%%%%%%%%

\end{document}